\documentstyle[11pt,aaspp4,flushrt,tighten]{article}
\singlespace

\font\gkvec=cmmib10                         %for boldface lowercase Greek
\def\bbeta{\hbox{{\gkvec\char12}}}         %bold face beta

\def\spose#1{\hbox to 0pt{#1\hss}}
\def\lta{\mathrel{\spose{\lower 3pt\hbox{$\mathchar"218$}}
     \raise 2.0pt\hbox{$\mathchar"13C$}}}
\def\gta{\mathrel{\spose{\lower 3pt\hbox{$\mathchar"218$}}
     \raise 2.0pt\hbox{$\mathchar"13E$}}}
\def\np{n_{e^+}}

\def\gamav{\langle\gamma_e\rangle}

\def\tiln{n'}
\def\wiell{\widetilde\ell}

\begin{document}
\title{Relativistic Winds from Compact Gamma-Ray Sources: \break
II. Pair Loading and Radiative Acceleration in Gamma-ray Bursts}

\author{Christopher Thompson\altaffilmark{1} and Piero 
Madau\altaffilmark{2,3}}

\altaffiltext{1}{Department of Physics and Astronomy, University of North 
Carolina, Chapel Hill, NC 27599.}
\altaffiltext{2}{Institute of Astronomy, Madingley Road, 
Cambridge CB3 0HA, UK.}
\altaffiltext{3}{Institute for Theoretical Physics, University of California, 
Santa Barbara, CA 93106--4030.}

\begin{abstract}

We consider the effects of rapid pair creation by an intense pulse of
$\gamma$-rays propagating ahead of a relativistic shock.  Side-scattered
photons colliding with the main $\gamma$-ray beam amplify the density
of scattering charges. The acceleration rate of the pair-loaded medium
is calculated, and its limiting bulk Lorentz factor related to the
spectrum and compactness of the photon source.  One obtains, as a result,
a definite prediction for the relative inertia in baryons and pairs.
The deceleration of a relativistic shock in the moving medium, and the
resulting synchrotron emissivity, are compared with existing results
for a static medium.  The radiative efficiency is increased dramatically
by pair loading.  When the initial ambient
density exceeds a critical value, the scattering depth tranversed
by the main $\gamma$-ray pulse rises above unity, and the pulse is broadened.
This sets an upper limit to the pre-burst mass loss rate
of $\sim 10^{-5} M_\odot$ per year, and places significant constraints
on gamma-ray burst progenitors. An anisotropic 
$\gamma$-ray flux (on an angular scale $\Gamma^{-1}$ or larger) drives a large
velocity shear that greatly increases the energy in the seed magnetic
field forward of the propagating shock.  

\end{abstract}

% \twocolumn

\section{Introduction}

Physical models for $\gamma$-ray emission from a relativistic fireball,
and the ensuing synchrotron emission from the decelerating shock, 
have generally neglected the feedback of the intense $\gamma$-ray flux on
the dynamics of the fireball, or on the prompt and delayed emission.
There are, however, several reasons to believe that this interaction
can have an important influence on the observed multiwavelength emission
from gamma-ray bursts (GRBs). 
Pair creation by high energy photons will raise the radiative efficiency
of a shock (Thompson 1997); and -- if the ambient medium is sufficiently dense
-- it will significantly smear and broaden the observed time profile of the
prompt GRB.  The pairs also carry net forward momentum
which will couple effectively to the ambient medium if this contains
a strong enough magnetic field.  In this work we extend
previous results on radiative acceleration in the Thomson
(Noerdlinger 1974; O'Dell 1981; Phinney 1982; Kovner 1984) and 
Klein-Nishina (Madau \& Thompson 1999, hereafter Paper I) scattering regimes
to include the effects of pair creation.  Attenuation of the high
energy $\gamma$-ray spectrum by collisions with soft photons was
calculated by Baring and Harding (1997), but the dynamical effects
of pair creation and the feedback of radiative acceleration on the
scattering depth were not considered in that work.  The role of
pair creation in accelerating a trans-relativistic outflow from
a black-hole accretion disk with a hard X-ray spectrum has been considered
recently by Beloborodov (1999).

The plan of this paper is as follows.  In \S\,2 we calculate
the rate of acceleration by a thin shell of $\gamma$-rays, some of
which  side-scatter off ambient charges and collide with the main
$\gamma$-ray beam.  We show that acceleration is much faster (for a given
compactness) than in the Thomson limit, because the highest energy
photons provide most of the momentum.  The reduction in
the mean inertia per scattering accompanying pair creation 
significantly increases the limiting Lorentz factor over the value
for a baryonic plasma.  In \S\,3 we apply these results to GRB afterglows.  
We consider the deceleration of a relativistic
shock propagating into a medium that is itself moving at bulk relativistic
speed, and show how the time scaling of the synchrotron emission
is modified from the case of a static medium.  We estimate
the limiting ambient density above which the $\gamma$-ray pulse
experiences a large optical depth to scattering.  Finally, we note
that angular variations in the radiative force will excite non-radial
shear flows in the ambient medium that can greatly amplify the
pre-existing seed magnetic field.

In the following we will denote by $x$ and $x_s$ the incident and scattered 
photon energies in units of $m_ec^2$ in the (unprimed) lab frame, and with 
$\Gamma=(1-\beta^2)^{-1/2}$ the bulk Lorentz factor of the flow. 

\section{Pair loading}

Gamma-ray sources of high compactness are opaque to photons that move
at an angle with respect to the $\gamma$-ray flow.  When the source
spectrum extends to an energy $x \gg 1$, even soft photons will create pairs
via $\gamma\gamma\rightarrow e^+e^-$, with a cross-section close to
Thomson near threshold.  This implies that the source of $\gamma$-ray
radiation 
must itself expand at relativistic speed.  When the photons are beamed into
a narrow angle $\theta$ along the direction of motion, the threshold energy 
for pair production within the beam is increased to $x \sim \theta^{-1}$.
In Paper I we calculated the radiative acceleration of matter exposed to 
a strong source of $\gamma$-rays, including the full Klein-Nishina
suppression of the scattering cross-section, but did not include the
effects of pair creation.  
However, the same Compton scattering process that accelerates material
sitting close to a $\gamma$-ray source also creates side-moving photons.
When the integrated flux of photons of energy $x \geq 1$ is large, each 
side-scattered photon deposits its {\it entire} momentum, along with
the momentum of a second colliding photon, into an electron-positron pair.
The build-up of a pair plasma (`pair loading') through this process,
and the bulk acceleration that ensues, will be the subject of this section.

\subsection{Radiative Acceleration}

At very high compactness, the source of gamma-rays must be impulsive; 
otherwise the assumption of low scattering optical depth in the
accelerating medium is not self-consistent.  
We approximate the photon source as a radially propagating shell situated at 
$c(t-\Delta t) < r < ct$.  We found in Paper I that, in the Thomson
limit, a test particle surfing the photon shell attains a limiting
Lorentz factor 
\begin{equation}
\Gamma_{\rm max} = {\tau_c\over 2}\,\left({m_e\over\mu}\right)
\label{eq:gams}
\end{equation}
when the shell is thin enough that the acceleration length is much
less than the radius,
\begin{equation}\label{eq:thinshell}
{2\over 3}\Gamma_{\rm max}^2 c\Delta t \ll r.
\end{equation}
The parameter
\begin{equation}
\tau_c \equiv \sigma_T \Delta t {F\over m_ec^2},
\label{eq:tauc}
\end{equation}
is expressed in terms of the integrated energy flux
\begin{equation}
F = \int F_x dx,
\end{equation}
and characterizes the column density of photons $N_\gamma$
through the shell,  $\tau_c \sim \sigma_T N_\gamma \langle x\rangle$.
When the particles are coupled by a magnetic field and the 
electrons and positrons are cold, the mean mass per scattering
charge is
\begin{equation}
\mu = m_e + m_p\left(1+2{n_{e^+}\over n_p}\right)^{-1}.\label{eq:mueq}
\end{equation}
(We assume for simplicity a hydrogen plasma and neglect the inertia of
the original neutralizing electron component, but not of the pairs.)
The term $m_e$ in this expression must be replaced with the relativistic
inertia $\mu_e \equiv{4\over 3}\gamav m_e$ when the electrons and
positrons are hot.

The optical depth to pair creation is proportional 
to the optical depth parameter $\tau_c$, which is in turn proportional
to the duration $\Delta t$ of the gamma-ray pulse, and the source compactness 
$L  \sigma_T/(4\pi\mu c^3R)$.  Here, 
$R$ and $L$ are the source radius and luminosity, respectively.
We assume that the incident spectrum is broadband, extending well
above and below $x \sim 1$.  As long as $\tau_c \gg 1$,
even a soft photon that is side-scattered can be absorbed by the main
photon beam.  The accelerating material is, in this situation,
{\it photon rich}:  a side-scattered photon that finds itself above the pair
creation threshold has a much higher probability of colliding with
another photon than of Compton scattering a second time.  Indeed, the
number of scattering particles in the relativistically expanding
material that emits the photons is suppressed by a factor of
$\Gamma^{-1}$ with respect to the number of photons.

We now calculate the momentum deposited when a (soft) photon $B$ is
side-scattered through an angle $\theta_s$ and collides with a (hard)
photon $A$.  The combined process of scattering followed by pair creation
deposits a net forward momentum $(x^A + x^B)m_e c$ in the accelerating
medium.  
When the scattering charge is cold,  the initial energy of
the scattered photon in the rest frame of the scattering charge is
$x' = x\Gamma(1-\beta)$.  A similar relation, 
$x_s'  =  x_s\Gamma (1-\beta \cos\theta_s')$, holds for the photon
after scattering.  The scattering angle in the frame of the
charge is given by $\cos\theta_s' = 
(\cos\theta_s-\beta)/(1-\beta\cos\theta_s)$.  
The lab frame energy $x^B$ before scattering can be
related to the energy $x^B_s$ following scattering through
\begin{equation}\label{eq:xborig}
x^B = {\Gamma x^B_s(1-\beta\cos\theta_s)\over  (1-\beta)\Gamma - 
(1-\cos\theta_s)x^B_s},
\end{equation}
which approaches $x^B\rightarrow x^B_s (1-\beta\cos\theta_s)/(1-\beta)$
in the Thomson limit ($x' = x_s'$).  In these variables, the cross-section 
for the collision between photons $A$ and $B$ (scattered) is
\begin{equation}
\sigma_{\gamma\gamma} = {3\over 16}\sigma_T\,(1-\xi^2)
\left[(3-\xi^4)\,\ln\left({1+\xi\over1-\xi}\right)
-2\xi(2-\xi^2)\right]
\end{equation}
(Jauch \& Rohrlich 1976), where
\begin{equation}
\xi=\sqrt{1-{2\over x^A x^B_s(1-\cos\theta_s)}}.
\end{equation}
The cross-section   
reaches a maximum of $0.25\sigma_T$ at $x^A=2x^A_{\rm min}$ ($\xi=0.71$). 

Most of the momentum deposited through this process involves the
side-scattering of very soft photons, which collide with hard gamma-rays
to produce energetic (and almost radially moving) pairs.
The minimum (threshold) energy of the radial photon can be expressed
as a function of the energy and propagation angle of the side-scattered
photon,
\begin{equation}
x^A_{\rm min}(x^B_s,\theta_s) = {2\over x^B_s(1-\cos\theta_s)}.
\end{equation}
Thus, a photon of energy $x^B_s \ll \langle x\rangle$ that is 
side-scattered through an angle $\sim \theta_s$ creates a pair of energy
$x^{\rm pair} \sim 4/\theta_s^2x^B_s$.  During the first stages
of the acceleration, this pair will be relativistic in the bulk frame, 
\begin{equation}
2\gamma_e({\rm injected}) \sim {x^{\rm pair}\over (1+\beta)\Gamma} \gg 1.
\end{equation}
(Recall that the photon beam is assumed to be radially streaming.)

In the presence of a sufficiently
strong magnetic field, this momentum is effectively communicated to the
other charges.  Inspection of eq. (37) in Paper I
shows that this coupling rapidly becomes stronger as $\Gamma$ increases.
In the case of a $\gamma$-ray fireball of duration $\Delta t
\sim 10$s and photon optical depth  $\tau_c \sim 300$ (eq. [\ref{eq:tauc}]),
propagating into an interstellar medium with $B_0 \sim 3\times 10^{-6}$ G,
the coupling is strong only for pairs of lab frame energy $x^{\rm pair}
\sim $ few.  However, above a Lorentz factor 
\begin{equation}
\Gamma \sim 3.5\,\left({B_0\over 3\times 10^{-6}~{\rm G}}\right)^{-1/5}\,
\left({\tau_c/\Delta t\over 30~{\rm s^{-1}}}\right)^{1/5}\,
\left({x^{\rm pair}\over 100}\right)^{2/5}
\end{equation}
even energetic pairs deposit their entire momentum.  In the process,
all particles will share a bulk radial Lorentz factor.  In the calculations
that follow, we assume that $B_0$ is strong enough (and $\Delta t$ long
enough) to assure effective coupling; but this constraint should be kept in
mind when applying the formulae below.

When $\tau_c \gg 1$, each newly created $e^+e^-$ Compton cools on the main
photon beam.  Indeed, it drops to an energy far below the injection energy over
the time that the medium accelerates;  this starting assumption will
be justified {\it a posteriori} in \S \ref{secmeane}.  In the process
of scattering off an anisotropic radiation field, the cooling particle
absorbs an additional momentum -- similarly to 
what happens in the Compton rocket (O'Dell 1981; Phinney 1982).  
The effect of this {\it Compton afterburn} on the net momentum  deposited
in the medium depends on whether the Compton-scattered photons are absorbed
by pair creation.  When they are not absorbed, the net effect is
to increate the total momentum absorbed per photon collision to
\begin{equation}\label{eq:totmom}
f\times(x^A + x^B)m_ec.
\end{equation}
Because the magnetic field becomes nearly transverse as the medium
accelerates, the numerical factor works out to  $f = {5\over 3}$  (Paper I).
Secondary pair creation can be neglected after the bulk Lorentz factor
reaches a critical value, but during the early stages of the acceleration 
the side-scattered photons are absorbed.  The net effect then is to
cancel the additional momentum absorbed by the cooling charge, 
and to force $f \rightarrow 1$.    We discuss
further details of the pair-photon cascade induced by the injection of
an energetic pair in \S \ref{inertia}, and for now summarize this process
in terms of mean relativistic inertia $\mu_e \simeq {4\over 3}\gamav m_e$ 
of the daughter $e^\pm$, and a multiplicity factor $N_{\rm mult}$.   When 
secondary pair creation can be neglected, one has $N_{\rm mult} \simeq 1$; 
otherwise 
\begin{equation}\label{eq:nmult}
2\gamav \,N_{\rm mult} \simeq {x^A + x^B\over (1+\beta)\Gamma}.
\end{equation}
The quantity on the RHS is the total energy of the injected pair in the
bulk frame.  

Let us now calculate the rate of acceleration due to injection of
energetic pairs which cool down rapidly off the main photon beam, to a mean
Lorentz factor $\gamav \ll \langle x^A + x^B\rangle$.  
In the fluid approximation the rate of change of the bulk momentum
is given by 
$$
{\partial\over\partial t}
\Bigl[\Gamma^2\beta c(2\mu_e\tiln_{e^+}+m_p\tiln_p)\Bigr]
+{\bf\nabla}\cdot\Bigl[\bbeta\,\Gamma^2\beta c^2(2\mu_e\tiln_{e^+}+m_p\tiln_p)
\Bigr] =\;\;\;\;\;\;\;\;\;\;\;\;\;\;\;\;\;\;\;\;\;\;$$
\begin{equation}
\;\;\;\;\;\;\;\;\;\;\;\;\;\;\;\;\;\;\;\;\;\;\;\;
\int (x^A+x^B)f\, {F_x(x^A)\over x^A} {I(x^B_s,\theta_s)\over x^B_s} 
\sigma_{\gamma\gamma}(1-\cos\theta_s)\, dx^A dx^B_s d\Omega.
\label{eq:eqmom}
\end{equation}
where the bulk-frame charge densities are denoted by a prime.
The rate pair creation induced 
by side-scattering of soft photons per unit volume,
\begin{equation}
{\partial\over\partial t}(\Gamma\tiln_{e^+}) + {\bf\nabla}\cdot
(\Gamma\bbeta c \tiln_{e^+}) = 
\int N_{\rm mult}{F_x(x^A)\over x^A} {I(x^B_s,\theta_s)\over x^B_s} 
\sigma_{\gamma\gamma}(1-\cos\theta_s)\, dx^A dx^B_s d\Omega.
\label{eq:eqpair}
\end{equation}
The Compton force acting on transrelativistic or subrelativistic
particles has been neglected.

Substituting the equations of continuity into (\ref{eq:eqmom}) yields
$$
{d\Gamma\beta\over dt} = {\partial\Gamma\beta\over\partial t}+
c\bbeta\cdot{\bf\nabla}(\Gamma\beta) =\;\;\;\;\;\;\;\;\;\;\;\;\;\;\;
\;\;\;\;\;\;\;\;\;\;\;\;\;\;\;\;\;\;\;\;\;\;\;\;\;\;\;\;\;\;\;\;\;\;
$$
\begin{equation}
{1\over m_e(2\mu_e\np + m_pn_p)c^5} \int \left[(x^A+x^B)f
- \left({\mu_e\over m_e}\right)\,N_{\rm mult}\,2\Gamma\beta \right]\, 
{F_x(x^A)\over x^A}\,{I(x^B_s,\theta_s)\over x^B_s} \sigma_{\gamma\gamma}\,
(1-\cos\theta_s)\, dx^A dx^B_s d\Omega.
\label{eq:gamac}
\end{equation}
The value of the second term in the brackets depends on the importance
of secondary pair creation.  During the last stages of acceleration
($f \simeq {5\over 3}$) one has $\mu_e \simeq m_e$ and $N_{\rm mult} 
\simeq 1$ (\S \ref{secmeane}).  Otherwise, we substitute $f=1$ and expression 
(\ref{eq:nmult}) for $N_{\rm mult}$, and find that the quantity
in brackets is $(x^A+x^B)$ multiplied by $(1-{1\over 3}\beta)/ (1+\beta)$.
Both these results can be expressed in terms of a $\beta$-dependent
Compton afterburn parameter
\begin{equation}\label{eq:fdef}
f_\beta = \left\{ \begin{array}{ll} {5\over 3} & \mbox{({\rm no~secondary~
pair creation});} \\ 
{1-\beta/3\over 1+\beta} & \mbox{({\rm secondary~pair~creation}).} \\
\end{array}
\right.
\end{equation}

The photon shell will be parameterized by a radial coordinate
$\varpi$, measured from its inner boundary,
\begin{equation}
r = c(t-\Delta t) + \varpi.
\end{equation}
The scattering occurs at position $\varpi_s$ and time $t_s$, and the
collision at position $\varpi_\pm$ and time $t_\pm$. Hence,
\begin{equation}
\varpi_s - \varpi_\pm = c(t_\pm-t_s)\cos\theta_s.
\end{equation}
The intensity $I(x^B_s,\theta_s)$ of
side-scattered photons is related to the flux $F_x(x^B)$ of unscattered
radiation in the thin shell (cf. eq. [\ref{eq:thinshell}]) via
\begin{equation}
I(x^B_s,\theta_s,\varpi_\pm)=
\int_{\varpi_\pm}^{c\Delta t} 
(2\np+n_p)\,F_x(x^B)\, (1-\beta) {d\sigma\over d\Omega}
\exp\left[-\tau_{\gamma\gamma}(x^B_s,\theta_s,\varpi_s-\varpi_\pm)\right]
d\varpi_s.
\end{equation}
Note that the radial position $\varpi_s$ of the scattering
site within the photon shell always lies outside the position 
$\varpi_\pm$ of pair creation, because the radial speed of the
side-scattered photon is less than $c$.
The optical depth to pair creation experienced by the side-scattered photon
is
\begin{equation}
\tau_{\gamma\gamma}(x^B_s,\theta_s,\varpi_s-\varpi_\pm) = 
\int_{\varpi_s}^{\varpi_\pm}\,{d\varpi^\prime\over c}\,
\int_{x^A_{\rm min}(x_s^B,\theta_s)}^\infty dx^A\,{F_x(x^A)\over x^A}\,
(1-\cos\theta_s)\,
\sigma_{\gamma\gamma}\Bigl[x^A/x^A_{\rm min}(x^B_s,\theta_s)\Bigr].
\end{equation}

When the optical depth along the ray from $\varpi_\pm$ to $\varpi_s$
is much larger than unity,  the momentum deposited by side-scattering and
pair creation is localized, and the $x^A$-integral can be performed directly:
\begin{equation}\label{accexp}
{d\over dt}(\Gamma\beta) = 
{f_\beta\over \mu c^2}\,
\int (\langle x^A\rangle + x^B)\, F_x(x^B) (1-\beta)\, 
d\sigma {dx^B_s\over x^B_s},
\end{equation}
where we have made use of eq. (\ref{eq:mueq}) for the mean mass $\mu$
per scattering charge.  
The average energy of the undeflected, radial photon is
\begin{equation}
\langle x^A \rangle
\equiv {\int F_x(x^A) \sigma_{\gamma\gamma} dx^A \over \int F_x(x^A) 
\sigma_{\gamma\gamma} dx^A/x^A},
\end{equation}
after weighting by cross-section and photon-spectrum.
We write 
\begin{equation}\label{eq:kdef}
\langle x^A\rangle(x_s^B,\theta_s) = k(\alpha)\,x^A_{\rm min}(x^B_s,\theta),
\end{equation}
from here on.  The constant $k$ is plotted as a function of
high energy spectra index in Fig. 1.

It is now necessary to prescribe the radiation spectrum.
We assume a simple broken power-law form, appropriate for GRB 
sources, with photon number flux (per logarithm of energy)
constant below a break energy $x_{\rm br} \sim 1$ and photon index
$\alpha$ above the break:
\begin{equation}\label{eq:specdef}
F_x(x^A)= \left\{ \begin{array}{ll} F_x(x_{\rm br}) & \mbox{$x^A<x_{\rm br}$;}
 \\ F_x(x_{\rm br}) \left(x^A/x_{\rm br}\right)^{-\alpha+1} & 
\mbox{$x^A>x_{\rm br}$.} 
\end{array}
\right. 
\end{equation}
For the assumed spectral shape, low energy photons ($x<x_{\rm br}$) are
side-scattered at a slightly higher rate than are photons near the break
energy $x_{\rm br}$ (the scattering cross-section is not KN-suppressed).
These low energy photons then pair produce off photons of energy
$x \gg x_{\rm br}$, and deposit a relatively large momentum. 
It will be convenient to introduce the modified parameter
\begin{equation}\label{eq:taucb}
\tau_c^\prime = \tau_c \times {x_{\rm br}\,F_{x_{\rm br}}\over F} =
\sigma_T \Delta t {x_{\rm br}F_{x_{\rm br}}\over m_ec^2},
\end{equation}
which rescales $\tau_c$ by the ratio of the energy flux at frequency
$x_{\rm br}$ to the total energy flux.

The energy of the side-scattered photon cannot, however, be taken
to be arbitrarily small.  When $x^B_s\ll 1$, the radiation shell becomes 
optically thin to pair creation.  This leads to an {\it upper} bound
on the energy of the photo-pair, which we now evaluate. 
Each side-scattered photon overlaps the shell for a time
$\sim \Delta t/(1-\cos\theta_s)$, but the collision rate is
proportional to $(1-\cos\theta_s)\sigma_{\gamma\gamma}$.  The optical
depth to pair creation therefore depends on $\theta_s$ only implicitly
through $x^A_{\rm min}$:
$$
\;\;\;\;\;\;\tau_{\gamma\gamma}(x^B_s\,\theta_s) = {\Delta t\over m_ec^2}
\int_{x^A_{\rm min}(x^B_s,\theta_s)}^\infty dx^A {F_x(x^A)\over x^A} 
\sigma_{\gamma\gamma}\Bigl[x^A/x^A_{\rm min}(x^B_s,\theta_s)\Bigr]
$$
\begin{equation}\label{taugmin}
\equiv\;\;\; k_2(\alpha)F_x\Bigl[x^A_{\rm min}(x^B_s,\theta_s)\Bigr]\,
{\sigma_T \Delta t\over m_ec^2} 
> 1.
\end{equation}
The dimensionless function $k_2$ is plotted against spectral index 
$\alpha$ in Fig. 1.
Defining a characteristic optical depth $\tau_c^\prime$ (eq. [\ref{eq:taucb}]) 
through the photon shell at frequency $x_{\rm br}$,
and noting that in general $x^A_{\rm min} \gg x_{\rm br}$, one gets 
an upper bound on the mean energy $x^{\rm pair} = \langle x^A + x^B\rangle
\simeq \langle x^A\rangle$,
\begin{equation}
{x^{\rm pair}_{\rm max}\over x_{\rm br}}
= k\,\left[k_2\,{\tau_c^\prime\over x_{\rm br}}\right]^{1/(\alpha-1)}.
\label{eq:xpair}
\end{equation}
This upper bound is approximately independent of the scattering $\theta_s$
(even though the most probable energy of the side-scattered photon that
collides to form a pair of energy $x^{\rm pair}_{\rm max}$ does depend on
$\theta_s$).  The corresponding lower bound on the energy of the
scattered photon is
\begin{equation}\label{eq:xbsmin}
x^B_s >  {2k\over x^{\rm pair}_{\rm max}(1-\cos\theta_s)}
\sim {4k\Gamma^2\over x^{\rm pair}_{\rm max}}.
\end{equation}
This energy sits below (above) the break energy $x_{\rm br}$ when
$\Gamma$ is less than (greater than) $2\Gamma^\star$, where
\begin{equation}
\Gamma^\star = {1\over 4 k^{1/2}}\,
(x_{\rm br}x^{\rm pair}_{\rm max})^{1/2}.
\end{equation}
Thus, the supply of side-scattered photons is smaller when $\Gamma > 
2\Gamma^\star$, and the acceleration rate is reduced accordingly.

We first calculate the acceleration at low Lorentz factors,
$\Gamma \ll \Gamma^\star$.  The softness of the side-scattered photons
allows us to perform the integral (\ref{accexp}) over 
$x^B_s$ in the Thomson regime, $dx^B_s/x^B_s = dx^B/x^B$.  
At this stage of the acceleration, 
$\langle x^A\rangle = x^{\rm pair}_{\rm max} \gg x^B$.
Setting a lower bound (\ref{eq:xbsmin}) to $x^B_s$, we obtain
\begin{equation}
{d\over dt}(\Gamma\beta) \simeq 
{f_\beta\,x^{\rm pair}_{\rm max}\,F_x(x_{\rm br})\over m_ec^2}\,
\left({m_e\over\mu}\right)\,(1-\beta)
\,\int d\sigma,
\end{equation}
and we have
\begin{equation}
{d\over dt}(\Gamma\beta) = f_\beta\,\tau_c^\prime\,
\left({x^{\rm pair}_{\rm max}\over x_{\rm br}}\right)\,
\,\left({m_e\over\mu}\right)\,{1-\beta\over\Delta t}\;\;\;\;\;\;
(\Gamma \ll \Gamma^\star).\label{eq:accone}
\end{equation}
It is worth recalling the physical origin of the various factors
in this expression:  $f_\beta$ is the Compton afterburn parameter
(eq. [\ref{eq:fdef}]) which describes the influence of secondary
Compton scattering and pair creation on the net momentum injected
per high energy pair;
$\tau_c^\prime$ (eq. [\ref{eq:taucb}]) is a dimensionless parameter,
proprotional to the energy flux at frequency $x_{\rm br}$ and 
comparable to the optical depth to $\gamma-\gamma$ collisions through
the photon shell;  $x^{\rm pair}_{\rm max}$ (eq. [\ref{eq:xpair}]) is 
the maximum energy of the injected pair in the lab frame (above which
the optical depth to pair creation through the shell is small);  
$\mu$ (eq. [\ref{eq:mueq}]) is the mean inertia per scattering charge;
and $\Delta t$ is the temporal width of the photon shell.

To approach the high-Lorentz factor regime (where $\Gamma \gg 2\Gamma^\star$)
we approximate the scattering charges as being cold, 
$\gamav = 1$ and $\mu_e = m_e$.  This approximation turns out to 
be justified, as in this regime the time for freshly injected
pairs to Compton cool in the bulk frame is short compared with the
acceleration time (\S \ref{secmeane}).  Since the minimum value of
$x^B$ lies above $x_{\rm br}$, the frequency dependence of $F_x$
gives an additional factor of 
\begin{equation}
{1\over \alpha}\left[{x^{\rm pair}_{\rm max}x_{\rm br}
(1-\beta)(1-\cos\theta_s)\over 
2k(1-\beta\cos\theta_s)}\right]^{\alpha-1}
\end{equation}
after performing the integral over $x^B_s$.  Here,
the factor in brackets is $x_{\rm br}$ divided by the minimum
value of $x^B$ before scattering (eqs. [\ref{eq:xborig}] and 
[\ref{eq:xbsmin}]).  After substituting the relation
$(1-\cos\theta_s)/(1-\beta\cos\theta_s) = (1-\cos\theta_s^\prime)/
(1+\beta)$, the acceleration rate can be expressed as
\begin{equation}
{d\over dt}(\Gamma\beta) = 
{2f_\beta k \over \alpha}\,
\left({x_{\rm br}x^{\rm pair}_{\rm max}\over 2x_{\rm br}k}\right)^\alpha\,
{F_x(x_{\rm br})\over m_ec^2}
(1-\beta)\,\int d\Omega^\prime \left[(1-\cos\theta_s^\prime)\,
{1-\beta\over 1+\beta}\right]^{\alpha-1}\,{d\sigma\over d\Omega^\prime},
\end{equation}
which becomes 
\begin{equation}
{d\over dt}(\Gamma\beta) = 
f_\beta\,\tau_c^\prime\,\left({x^{\rm pair}_{\rm max}\over x_{\rm br}}\right)\,
\left({\Gamma^\star\over\Gamma}\right)^4\,
{1\over 2(\Gamma^\star)^2\Delta t}
\;\;\;\;\;\;\;\;\;(\Gamma \gg \Gamma^\star)\label{eq:acctwo}
\end{equation}
for $\alpha = 2$.  This expression matches onto (\ref{eq:accone}) at
$\Gamma = \Gamma^\star$.

Let us compare the rates of acceleration just derived with the
corresponding expression 
\begin{equation}
{d\Gamma \over dr}=\Gamma^2(1-\beta)^2\wiell {R\over r^2}.  \label{eq:eqps}
\end{equation}
for Thomson scattering without pair creation (Noerdlinger 1974; Paper I).
When $1 \ll \Gamma \ll \Gamma^\star$, the acceleration rate 
(\ref{eq:accone}) is larger by $\sim 2f_\beta\,x^{\rm pair}_{\rm max}$
than for Thomson scattering by a spectrum localized near $x_{\rm br}$;
and when $\Gamma \gg \Gamma^\star$ the acceleration rate
(\ref{eq:acctwo}) is larger by $\sim 2f_\beta\,x^{\rm pair}_{\rm max}\,
(\Gamma/\Gamma^\star)^{-2}$.

Pair creation no longer increases the speed of the bulk flow when
two colliding photons of comparable energy, 
$x^B \sim x^A \sim 2k/x^B_s(1-\cos\theta_s) \sim x^{\rm pair}_{\rm max}$
have an optical depth to pair creation that is close to unity. 
The resulting maximum Lorentz factor is
\begin{equation}
\Gamma_{\rm max}^{\rm pair} \sim 
{x^{\rm pair}_{\rm max}\over \sqrt{2k}} =
\left({k\over 2}\right)^{1/2}\,
\left[k_2 x_{\rm br}^{\alpha-2}\,\tau_c^\prime\right]^{1/(\alpha-1)}.
\label{eq:gamp}
\end{equation}
Continued scattering in the pair-loaded flow increases its speed 
further:  the limiting Lorentz factor is
\begin{equation}
\Gamma_{\rm max} \simeq {1\over 2}\,\tau_c,\label{eq:gammax}
\end{equation}
from eq. (\ref{eq:gams}).  This
is comparable to (\ref{eq:gamp}) for $\alpha = 2$, but larger
for softer high energy spectra.   We conclude that the main 
effects of pair creation are i) to increase the rate of acceleration at
low Lorentz factor;  and ii) to reduce the mean mass per scattering
charge to $\mu \simeq m_e$,  thereby allowing much higher terminal 
Lorentz factors.  

\subsection{Mean Energy of the Pairs in the Bulk Frame}\label{secmeane}

It remains to calculate the mean energy $\gamav$ of the pairs in the
bulk frame, and show that our starting assumption that 
$\gamav \ll x^{\rm pair}_{\rm max}$ is self-consistent.  During the 
first stages of acceleration an 
electron or positron will be injected with a characteristic 
energy $\gamma_e^{\rm max} \sim {1\over 2}x^{\rm pair}_{\rm max}$,
and will cool off the radial photon flux when 
$x_{\rm br} \ll x^{\rm pair}_{\rm max}$.
At the same time, the accelerated
medium is compressed (Paper I), and the pairs are adiabatically
heated at a rate $\gamav^{-1}(d\gamav/dt) = {1\over 3}\tiln^{-1}(d\tiln/dt)
= {1\over 3}\Gamma^{-1}(d\Gamma/dt)$.
The equilibrium energy $\gamav$ that results from a balance between
compressional heating and Compton cooling lies far below the injection energy,
and so we will assume that the heating process has isotropized the
momenta of the pairs in the bulk frame.  From eq. (38) of Paper I,
\begin{equation}
\gamav = {1\over 2}\,\left({t_{\rm cool}^0\over t_{\rm accel}^0}\right)^{1/2},
\end{equation}
where the reference (lab-frame)
cooling time is $t_{\rm cool}^0 = m_ec^2/\sigma_T \Gamma F_\Gamma 
\simeq (\Gamma/2\tau_c^\prime)\Delta t$, and $F_\Gamma \simeq F/2\Gamma^2$ is
the energy flux in the boosted frame. 
The acceleration rate is decreased by a factor
$\sim \gamav^{-1}$ from the values (\ref{eq:accone}) and (\ref{eq:acctwo})
calculated above for a cold plasma:
\begin{equation}
{1\over\Gamma}{d\Gamma\over dt} \simeq {1\over\gamav\,t_{\rm accel}^0}.
\end{equation}
For a spectrum $\alpha = 2$,
\begin{equation}
{t_{\rm cool}^0\over t_{\rm accel}^0} = {f_\beta\,x^{\rm pair}_{\rm max}
\over 4\Gamma^2} = {4f_\beta\,k\over x_{\rm br}^2}
\left({\Gamma\over\Gamma^\star}\right)^{-2}
\end{equation}
at $1 \ll \Gamma \ll \Gamma^\star$, and
\begin{equation}
{t_{\rm cool}^0\over t_{\rm accel}^0} = {4f_\beta\,k\over x_{\rm br}^2}
\,\left({\Gamma\over\Gamma^\star}\right)^{-4}
\end{equation}
at $\Gamma \gg \Gamma^\star$.  The equilibrium Lorentz factor works out to
$\gamav \simeq 1.4\Gamma^{-1}\,(x^{\rm pair}_{\rm max}/100)^{1/2}$
for a pair-loaded plasma moving slowly ($\Gamma < \Gamma^\star$).

\subsection{Relative Inertia in Pairs and Baryons}
\label{inertia}

The distribution of inertia between (cold) pairs and baryons within
the accelerated material has important implications for the
radiative efficiency of the forward shock.  The number of pairs
multiplies greatly as each hard $e^\pm$ (of bulk frame energy
$\gamma_e \sim x^{\rm pair}_{\rm max}/4\Gamma$) cools by side-scattering
the photon beam.  During the initial stages of the acceleration,
the side-scattered photons are well above the pair creation threshold;
little additional momentum is absorbed from the photon beam
when they collide with soft photons.  The momentum
imparted to the medium by the cooling of the hard $e^\pm$ 
is cancelled.  In this regime, $f \simeq 1$.

Balancing the radial momentum density in the lab frame,
under the assumption that each injected hard pair leads
to $N_{\rm mult} \gg 1$ softer pairs with an isotropic distribution
in the bulk frame, we obtain
\begin{equation}
x^{\rm pair}_{\rm max}\,{\np\over N_{\rm mult}}\,m_ec =
\Gamma\left(2\mu_e\np+m_pn_p\right)c.
\end{equation}
Here, $\mu_e = {4\over 3}\gamav m_e$  when the pairs are at least
mildly relativistic and the multiplicity $N_{\rm mult}$ is 
given by $x^{\rm pair}_{\rm max}/4\Gamma\gamav$ (eq. [\ref{eq:nmult}]).
This equation inverts to
\begin{equation}\label{eq:pairint}
{2\np \mu_e\over m_p n_p} = 2.
\end{equation}
The medium becomes very heavily loaded by pairs as it accelerates
to bulk relativistic speed (where $\langle\gamma_e\rangle \simeq 1$).  
The fraction of the inertia carried by the hadrons becomes
\begin{equation}\label{eq:fh}
f_h  = {m_p n_p\over 2\mu_e\np + m_pn_p} \simeq {1\over 3}
\end{equation}
in this regime.

Compton scattering of photons of lab-frame energy $\sim x_{\rm br}$
by the hard $e^\pm$ is initially in the Klein-Nishina regime, where the 
$e^\pm$ loses a significant fraction of its energy in one scattering.  
Cooling continues in the KN regime as
long as  $\gamma_e^2(x_{\rm br}/2\Gamma) > \gamma_e$,
that is, as long as $\Gamma < (2k)^{1/2}\Gamma^\star$.  As the
medium attains a Lorentz factor not far below the maximum value
(\ref{eq:gammax}), the hard $e^\pm$ are injected with a small
enough bulk frame energy that most of the energy lost to Compton
cooling is {\it not} reabsorbed by photon collisions.   This occurs
when $\gamma_e^2(x_{\rm br}/2\Gamma)(x^{\rm pair}_{\rm max}/2\Gamma) < 1$,
that is, when $(\Gamma/\Gamma_{\rm max})^4 > 
{1\over 4}\,k^3 k_2^{3/(\alpha-1)}\, 
(\tau_c^\prime/x_{\rm br})^{(7-4\alpha)/(\alpha-1)}$.  
Above this Lorentz factor, $f \simeq {5\over 3}$.

Implicit in this calculation of the maximum Lorentz factor is the 
assumption that the photon flux is high enough to supply the needed
momentum.  This constraint translates into an upper bound on the
density of the medium surrounding the $\gamma$-ray source.  The
strongest constraint comes from the requirement that the phase
of pair-dominated acceleration continues up to a Lorentz factor
$\Gamma^{\rm pair}_{\rm max}$ (eq. [\ref{eq:gamp}]).
This sets a lower bound to the flux of photons of energy 
$\sim x^{\rm pair}_{\rm max}$.

\section{Shock Deceleration, Pulse Broadening, and Afterglow} 

We now turn to consider the effects of radiative acceleration
on the dynamics of a decelerating, relativistic fireball.
The emission process operating in a GRB probably
covers a range of radius.  In some models, the emission is due
to optically thin synchrotron and/or inverse-Compton processes 
(M\'esz\'aros, Rees, \& Papathanassiou 1994; 
Sari, Piran, \& Narayan 1998); whereas in others the emission is by
inverse-Compton scattering at moderate Compton optical depth 
(Thompson 1994, 1997; Crider et al. 1997).  To focus the present
discussion, we assume that the dissipation extends outward from 
(at least moderately) large optical depth.
The optical depth $\tau_c$ (eq. [\ref{eq:tauc}]) through the photon
shell is
\begin{equation}
\tau_c = \tau_c^0\,\left({R\over R_\tau}\right)^{-2},
\end{equation}
outside the scattering photosphere. Here, $R_\tau$ is the radius at which the
ejecta become optically thin to scattering and the $\gamma$-ray
photons begin to stream ahead of the matter.
As long as pair creation and annihilation remain in equilibrium at
energy $x \sim 1$ in the bulk frame ($x \sim \Gamma_{\rm ej}$ in the
lab frame), the position of the Thomson photosphere is determined
implicitly by\footnote{Here $\alpha$ is the high-energy spectral index; 
eq. (\ref{eq:specdef}).}
\begin{equation}
F_x\Bigl(\Gamma_{\rm ej} x_{\rm br}\Bigr){\sigma_T\Delta t\over m_ec^2}
= {\tau_c^\prime\over x_{\rm br}}\,
\left({\Gamma_{\rm ej}\over x_{\rm br}}\right)^{-\alpha} \sim 1.
\end{equation}
(Recall that $\tau_c^\prime$ is $\tau_c$ rescaled by the factor 
$x_{\rm br} F_{x_{\rm br}}/F$.)  In this way, the initial value of $\tau_c$ 
can be related to the (initial) bulk Lorentz factor of the ejecta,
\begin{equation}\label{eq:taucval}
\tau_c^{\prime\,0} \sim (\Gamma_{\rm ej}^0)^{\alpha-1}\, 
x_{\rm br}^{2-\alpha}.
\end{equation}
This equation provides an upper bound on $\tau_c^{\prime\,0}$
if the $\gamma$-ray emission region is optically thin to scattering.

The width of the prompt $\gamma$-ray pulse is bounded below by
\begin{equation}
\Delta t \geq {R_\tau\over 2c(\Gamma_{\rm ej}^0)^2}.\label{eq:delt}
\end{equation}
More generally, we can define a minimum timescale for variability
driven by internal dissipation in the outflow,
\begin{equation}
\Delta t_{\rm var}(R) = {R\over 2c\Gamma_{\rm ej}^2}.
\end{equation}
This timescale can be related to the $\gamma$-ray fluence 
$\partial E/\partial\Omega$ (at energy $x_{\rm br}$ and per unit solid
angle) by noting that
\begin{equation}\label{eq:tauerel}
\tau_c^{\prime\,0} = \sigma_T {\partial E/\partial\Omega\over m_ec^2 R_\tau^2}.
\end{equation}
At the scattering photosphere, one finds
\begin{equation}\label{dtgamma}
\Delta t_{\rm var}(R_\tau) = 0.5\,
\left({x_{\rm br}\over 300}\right)^{(\alpha-2)/2}\,
\left[{4\pi(\partial E/\partial\Omega)\over 10^{53}~{\rm erg}}\right]^{1/2}\,
\left({\Gamma_{\rm ej}^0\over 300}\right)^{-(3+\alpha)/2}\;\;\;\;\;\;{\rm s}
\end{equation}
upon making use of eq. (\ref{eq:taucval}).
Should the overall duration $\Delta t$ of the burst be 
determined by the activity of the
central engine, one does not expect the afterglow emission to connect
up smoothly with the main burst.  However, the internal
dissipation could be triggered by an interaction with the external
medium:  for example, when a blob of magnetized ejecta is decelerated
and the causal propagation distance within the blob grows large enough
to allow reconnection (Thompson 1994; Begelman 1998).  In such
a hybrid model, the observation of distinct, overlapping
sub-pulses requires that the ejecta extend over a cone of
angular width $\theta \gg (\Gamma_{\rm ej}^0)^{-1}$, and that
the surface of the ejecta is wrinkled on angular scales larger than
$\sim (\Gamma_{\rm ej}^0)^{-1}$.  To simply the discussion that follows,
we assume that the burst is composed of a single pulse of width
$\Delta t \sim R_\tau/2(\Gamma_{\rm ej}^0)^2 c$.

\subsection{Deceleration of the Ejecta}

Pre-acceleration of the ambient medium by the prompt $\gamma$-ray 
pulse will slow the deceleration of the burst ejecta, and therefore
modify the synchroton emission from the forward shock.
The maximum Lorentz factor (eq. [\ref{eq:gammax}])
to which the ambient medium may be accelerated
decreases with distance from the matter photosphere.  
Let us assume, for the time being, that the ambient material is light
enough to allow acceleration to $\Gamma_{\rm amb} = \Gamma_{\rm max}$. Then
\begin{equation}\label{eq:gamamb}
\Gamma_{\rm amb}(R) = \Gamma_{\rm max}(R_\tau)\,
\left({R\over R_\tau}\right)^{-2},
\end{equation}
since $\Gamma_{\rm max}\propto \tau_c \propto R^{-2}$.
Thus, the kinetic energy per scattering charge behind the forward shock is
\begin{equation}
{\Gamma_{\rm ej}^2(R)\over\Gamma_{\rm amb}(R)}\mu c^2,
\end{equation}
in the lab frame.  Here, $\mu = m_e/(1-f_h)$ (eq. [\ref{eq:mueq}])
is the mean mass per scattering charge.  

The energy $E$ of the shock decreases with radius, because only a fraction
fraction $f_h$ of the post-shock particle energy is
carried by baryons that do not cool.  The calculation in \S \ref{inertia}
indicates that the inertia of the ambient medium becomes dominated
by leptons (eq. \ref{eq:fh}).  Energy is
deposited in shocked leptons at a rate
\begin{equation}\label{eq:edotl}
{\partial^2E_\pm\over\partial\Omega\partial t} = 
-(1-f_h){\Gamma_{\rm ej}^2\over \Gamma_{\rm amb}}
{m_p\over f_h}n_{p\,0}(R) R^2 c^3.
\end{equation}
We assume that these particless cool instantly.  When $f_h \ll 1$, the
remaining energy $E$ of the fireball is dominated by the kinetic energy
of the ejected matter (mass $M_0$), 
\begin{equation}\label{eq:ehadron}
{\partial E\over \partial\Omega} 
= \Gamma_{\rm ej}{\partial M_0\over\partial\Omega} c^2.
\end{equation}
One can calculate the time-scaling of the shock energy by balancing the
energy radiated (\ref{eq:edotl}) against the energy retained 
(\ref{eq:ehadron}).
Assuming the density profile around the center of the explosion to be
a powerlaw
\begin{equation}
n_{p\,0}(R) = {\rho_0(R)\over m_p} \propto R^{-\delta},
\end{equation}
one finds
\begin{equation}\label{eq:erad}
E(R) \propto R^{-\epsilon},
\end{equation}
with
\begin{equation}\label{eq:epsval}
\epsilon = 5-\delta.
\end{equation}
This compares with $\epsilon \simeq 3-\delta$ in the case where the
external medium is static or moves subrelativistically 
(e.g. Katz \& Piran 1997).  

More generally we will assume a power-law scaling (\ref{eq:erad}) of
the shock energy.  This assumption is motivated by the weak
dependence of $f_h$ (eq. [\ref{eq:fh}]) on the gamma-ray continuum
in the region where acceleration to bulk relativistic speed is possible.
At large radius $f_h \rightarrow 1$ and $\epsilon \rightarrow 0$,
because pairs with the miminum (shock-frame) energy 
$\sim \Gamma_{\rm ej} m_ec^2$ no longer cool effectively.
This allows a first evaluation of the influence of radiative acceleration
and pair loading on the initial stages of the deceleration of a relativistic
shock.  A more detailed numerical treatment of shock deceleration
including these effects is deferred to a later paper.

Balancing the shock energy with the energy deposited in shocked
particles, one obtains
\begin{equation}\label{eq:gamej}
\Gamma_{\rm ej}(R) = \Gamma_{\rm ej}(R_\tau)\,\left({R\over R_\tau}\right)
^{(\delta-\epsilon-5)/2}.
\end{equation}
The relation between shock radius and observed (post-burst) time $t$ becomes
\begin{equation}
t \propto R^{6-\delta+\epsilon}.
\end{equation}

Pair creation feeds back strongly on the radiative losses from
a relativistic outflow, and regions with extended non-thermal spectra
are expected to be much brighter than regions with quasi-thermal (e.g.
Wien) spectra (Thompson 1997).  It also introduces new characteristic
lengthscales and time-scales into the afterglow process.  One of these is
the radius $R(\Gamma_{\rm amb}\sim 1)$ outside of which the ambient
medium is accelerated to sub-relativistic speeds.  If the medium
were to remain pair loaded out to this radius, one would calculate
$R(\Gamma_{\rm amb}\sim 1)$ by setting $\mu \sim m_e$ in
eq. (\ref{eq:gams}); this would yield $R(\Gamma_{\rm amb}\sim 1)/R_\tau
\sim ({1\over 2}\tau_c)^{1/2}$.  
However, because several efoldings of the pair density are required
to push the ambient medium (assumed initially free of pairs) to bulk
relativistic speed, bulk relativistic motion effectively stops 
somewhat inside that radius.  A more accurate estimate of
$R(\Gamma_{\rm amb} \sim 1)$ is obtained by noting that
when $\tau_c$ is not much larger than unity, the pair density exponentiates
according to $\np/n_p \sim \exp(n_\gamma\sigma_Tc\Delta t) =
\exp(\tau_c/x_{\rm br})$.  Requiring that $2m_e\np \sim m_pn_p$
yields $\tau_c/x_{\rm br} \sim \ln(m_p/2m_e) \sim 7$.  
This value of $\tau_c$ is large enough that the photon shell is
optically thick to photon collisions, $\tau_{\gamma\gamma} \sim 
0.25\tau_c/x_{\rm br} \sim 2$.   Pairs continue to dominate baryons in
number out to a radius where $\tau_c/x_{\rm br} \sim 1$; but are not
produced in sufficient numbers to force the mean inertia per scattering
charge down to $\mu \sim m_e$.   The corresponding delay from the beginning
of the burst is given by the time it takes the forward shock to reach 
that radius.  In the case of a uniform external medium, $\Gamma_{\rm ej}
\propto R^{-5/2}$ and the time delay scales with radius as 
$R/2\Gamma_{\rm ej}^2c \propto R^6$.  This yields
\begin{equation}
{\Delta t(\Gamma_{\rm amb}\sim 1)\over \Delta t} \sim 
\left({\tau_c^0\over 10}\right)^3 \sim 
3\times 10^4\,\left({\tau_c^0\over 300}\right)^3.
\end{equation}

\subsection{Synchrotron Emission}

This pre-acceleration of
the ambient medium will modify the time-scaling of the synchrotron
emission from the forward shock propagating ahead of the burst ejecta. 
The pair-loading ensures that the Lorentz factor of the
shock-accelerated particles (in the shock frame) extends downward
to the minimum kinematically allowed value, $\gamma_e \sim \Gamma_{\rm ej}$
(Thompson 1997).  We consider here only the simplest emission model
involving synchrotron radiation by a power-law distribution of pairs,
$\partial n_e/\partial\gamma_e \propto \gamma_e^{-P}$ for rest frame
energies 
\begin{equation}
\gamma_e > {\Gamma_{\rm ej}\over\Gamma_{\rm amb}}.
\end{equation}
Closely related synchrotron models, in which
the inertia of the non-thermal particles is dominated by baryons, have
been constructed by M\'esz\'aros \& Rees (1997), Wijers, Rees \&
M\'esz\'aros (1997) and Waxman (1997).   The energy carried by the non-thermal
pairs is taken to be a constant fraction $\varepsilon_{\rm cr}$ of the total
post-shock particle energy in the shock frame, and the magnetic field is 
assumed to carry a constant fraction $\varepsilon_B$ of the post-shock
pressure.  As we discuss below, pre-acceleration of the ambient
medium by a collimated $\gamma$-ray beam will ensure that this second
assumption is satisfied, once the ejecta are moving sufficiently slowly.

The medium just ahead of the shock is compressed at the same time
as it is accelerated, so that the lab-frame density of baryonic
particles has increased to $n_p \simeq 2\Gamma_{\rm amb}^2 n_{p\,0}$
over the ambient value (eq. [17] of paper I).   The bulk frame energy density
behind the shock is, then
\begin{equation}
\left({\Gamma_{\rm ej}\over\Gamma_{\rm amb}}\right)^2 2n_{e^+} \mu c^2 
\simeq {2\Gamma_{\rm ej}^2\over f_h} \, n_{p\,0}\,m_p c^2.
\end{equation}
The equilibrium post-shock magnetic pressure therefore scales as 
\begin{equation}
{B_r^2\over 8\pi} \propto \varepsilon_B \Gamma_{\rm ej}^2 n_{p\,0}
\propto R^{-5-\epsilon},
\end{equation}
independent of $\delta$ in the shock rest frame.

We are now in a position to work out the time-scaling of the
synchrotron emission at fixed frequency $\nu$.  Pairs which radiate
at this frequency (Lorentz-boosted into the observer's frame) have an energy
\begin{equation}
\gamma_e \sim \left({2\pi\nu m_e c\over \Gamma_{\rm ej} eB_r}\right)^{1/2}
\propto R^{5/2 - (\delta-2\epsilon)/4}.
\end{equation}
When synchrotron cooling is rapid, the synchrotron emissivity
depends on the incoming flux of kinetic energy, but not explicitly
on post-shock magnetic field.  Assuming an powerlaw spectrum 
$\partial N_e/\partial\gamma_e \propto \gamma_e^{-P}$ of injected particles,
the energy released in photons of frequency $\nu$ is, per unit time 
\begin{equation}
{\nu\over R^2}{\partial L_\nu\over\partial \Omega} \simeq
\Gamma_{\rm ej}^2 \,\left[\varepsilon_{\rm cr}\,
\left({\gamma_e\over \Gamma_{\rm ej}/\Gamma_{\rm amb}}\right)^{2-P}\,
2\Gamma_{\rm ej}^2 n_{p\,0}(R){m_pc^3\over f_h}\right].
\label{eq:leq}
\end{equation}
The quantity in brackets denotes the rate at which particle energy accumulates
(in the bulk frame) per unit area of the shock; the prefactor
transforms to the lab frame.  The power law index 
\begin{equation}
\beta \equiv -{\partial \ln(\nu L_\nu)\over \partial\ln t}
\end{equation}
is expressed in terms of the synchrotron index $\alpha_{\rm sync} = 
{1\over 2}P$ via
\begin{equation}
\beta = {6\alpha_{\rm sync} + 2 - {1\over 2}\delta(3\alpha_{\rm sync}-1)
+2\epsilon\,\alpha_{\rm sync}\over 6-\delta+\epsilon}
\;\;\;\;\;\;(\Gamma_{\rm amb}=\Gamma_{\rm max};\;\;\rm fast~cooling).
\label{eq:betval}
\end{equation}
%When the shock loses energy rapidly ($\epsilon\gg 1$), this becomes
%$\beta = 2\alpha_{\rm sync}$.  
When the shock loses energy 
slowly ($\varepsilon \simeq 0$), one has $\beta = \alpha_{\rm sync} + 
{1\over 3}$ for a uniform medium ($\delta = 0$) and 
$\beta = {3\over 4}\alpha_{\rm sync} + {3\over 4}$ for a steady wind 
($\delta = 2$).

Now let us compare this result for the time-scaling of the synchrotron
emission, with the corresponding expression for a shock
propagating into a {\it static} external medium.  In that more familiar
case, the shock Lorentz factor, observer's time, and rest-frame magnetic
field scale as $\Gamma_{\rm ej} \propto 
R^{(\delta -\epsilon -3)/2}$, $t \propto R^{1/(4-\delta+\epsilon)}$,
$B_r \propto R^{-3/2-\epsilon/2}$, and $\gamma_e \propto 
R^{3/2-(\delta-2\epsilon)/4}$.  For those scalings, the time index is 
\begin{equation}
\beta = {6\alpha_{\rm sync} - 2 - {1\over 2}\delta(3\alpha_{\rm sync}-1)+
2\epsilon\,\alpha_{\rm sync}\over 4-\delta+\epsilon}
\;\;\;\;\;\;(\Gamma_{\rm amb}=1;\;\;\rm fast~cooling).
\end{equation}
This agrees with the scaling previously obtained whose energy decreases by 
Sari et al. (1998) and  M\'esz\'aros et al. (1998) if we substitute
expression (\ref{eq:epsval}) for $\epsilon$ (in the case of a strongly
radiative shock); or $\epsilon = 0$ (in the case of an almost adiabatic 
shock).  An adiabatic shock has $\beta = 
{3\over 2}\alpha_{\rm sync}-{1\over 2}$ for any value of $\delta$.

The regime of slow synchrotron cooling (at the observed frequency
$\nu$) is treated analogously.  We can now assume $\epsilon = 0$,
because the lowest energy $e^\pm$ must not be able to cool in this regime. 
One has 
\begin{equation}
\nu{\partial L_\nu\over\partial\Omega}
\propto \left(\gamma_e{\partial^2 N_e\over\partial\gamma_e\partial\Omega}
\right)\,\Gamma_{\rm ej}^2\gamma_e^2 B_r^2,
\end{equation}
where the total number of radiating charges is
\begin{equation}
\gamma_e{\partial^2 N_e\over\partial\gamma_e\partial\Omega}
\simeq 
{1\over 3-\delta}R^3 n_{p\,0}(R)\,\varepsilon_{cr}{m_p\over f_h m_e}
\left({\gamma_e\over\Gamma_{\rm ej}/\Gamma_{\rm amb}}\right)^{1-P}.
\end{equation}
The relation between synchrotron index and particle index softens
to $\alpha_{\rm sync} = (P-1)/2$, and the time index becomes identical
to (\ref{eq:betval}):
\begin{equation}
\beta = {6\alpha_{\rm sync} + 2 - {1\over 2}\delta(3\alpha_{\rm sync}-1)
\over 6-\delta}
\;\;\;\;\;\;(\Gamma_{\rm amb}=\Gamma_{\rm max};\;\;\rm slow~cooling)
\label{eq:betvalb}
\end{equation}
for $\epsilon = 0$.
Notice that the functional dependence of the time-index $\beta$ on
the spectral index $\alpha_{\rm sync}$ remains the same for both fast and
slow synchrotron cooling, when the ambient medium has been pre-accelerated
by the prompt $\gamma$-ray pulse and the lowest energy $e^\pm$ do not cool
radiatively ($\epsilon \simeq 0$).  If the injected spectrum of non-thermal
particles is constant -- let us take $P = 2$ -- then across a `cooling break'
the spectrum hardens from $\alpha_{\rm sync} = 1$ to $\alpha_{\rm sync}
= {1\over 2}$ and the time index hardens from $\beta = 
(8-\delta)/(6-\delta)$ to $(5-{1\over 4}\delta)/(6-\delta)$.  

By contrast, there is a stronger break in the time scaling
between regimes of fast and slow cooling if the external medium is static,
\begin{equation}
\beta = {6\alpha_{\rm sync}-{1\over2}\delta(3\alpha_{\rm sync}-1)
\over 4-\delta}\;\;\;\;\;\;(\Gamma_{\rm amb}=1;\;\;\rm slow~cooling).
\end{equation}
(M\'esz\'aros et al. 1998).  This simplifies to the familiar result $\beta = 
{3\over 2}\alpha_{\rm sync}$ in a uniform external medium.
For $P = 2$, the time index hardens from $\beta = 1$ to $\beta = 
(3-{1\over 4}\delta)/(4-\delta)$ across the `cooling break'.

\subsection{Anisotropic Radiation Pressure and Shearing of External
Magnetic Fields}\label{magtangle}

The prompt $\gamma$-ray pulse induces strong shearing motions in
the ambient medium that can amplify a seed magnetic field before
the forward shock hits.  An important question, that has yet to be resolved in
afterglow models, regards the effectiveness with which a weak ambient
magnetic field (e.g. $B_{\rm ex} \sim 3\mu$ G for a fireball interacting
with the ISM) will be amplified above the flux density $\sim \Gamma_{\rm ej}^2
B_{\rm ex}$ expected from laminar compression behind a relativistic shock.
In the non-relativistic limit (where pair creation can
be neglected), the limiting speed of the ambient medium is proportional to the 
$\gamma$-ray flux.  One therefore expects the flow outside the forward
shock to be strongly sheared on lengthscales in between
$(\Gamma_{\rm ej}^0)^{-1} R$ and $\theta_{\rm beam}R$, given that
the $\gamma$-ray emission is beamed within an angle $\theta_{\rm beam}$
but also varies on the smaller angular scale due to causal fluctuations
in the rate of internal dissipation.
This sheared flow will be Kelvin-Helmholtz unstable and, if there
is time before the shock hits,  the resulting turbulence will strongly
tangle a seed magnetic field.   Since the available time is
$\sim R/2\Gamma_{\rm ej}^2c$, one expects effective tangling only
after the ejecta have decelerated below $\Gamma_{\rm ej}
\sim (\Gamma_{\rm ej}^0)^{1/2}$.  Thereafter, near equipartition between
the magnetic pressure and the turbulent pressure in the sheared flow
appears plausible:   $B \sim (4\pi n_{p\,0}m_pc^2)^{1/2} \sim 0.1\,
(n_{p\,0}/1~{\rm cm^{-3}})^{1/2}$ G.  If no other instability (such
as a two-stream instability;  Medvedev \& Loeb 1999) raises
the field close to equipartition, one expects the burst and/or afterglow
lightcurves to contain features correlated with the onset of effective magnetic
shearing.

\subsection{Upper Bound on Pre-Burst Mass Loss: Pulse Broadening}

A strong constraint on the density of the ambient medium comes from the
requirement that the accelerated matter (which becomes loaded with pairs)
remain optically thin to Compton scattering, so that the $\gamma$-ray
pulse is not appreciably broadened.  This becomes a serious constraint
even if the scattering optical depth through the ambient medium
is small (cf. \S 3.3 of Paper I).  If all photons of frequency 
$> x$ are converted to pairs, then the scattering depth through
the resulting shell of pairs is 
\begin{equation}
\tau_T \simeq {\tau_c^\prime\over x_{\rm br}}\,
\left({x\over x_{\rm br}}\right)^{1-\alpha} =
{\tau_c^{\prime\,0}\over x_{\rm br}}\,\left({R\over R_\tau}\right)^{-2}\,
\left({x\over x_{\rm br}}\right)^{1-\alpha}.
\end{equation}
In order to prevent photons of this energy from being completely consumed,
the medium must be accelerated sufficiently that the photons are no
longer above the threshold energy for pair-creation in the bulk 
frame.\footnote{The material undergoing radiative acceleration remains
in causal contact with itself, and so the scattering depth through it is
not affected by the suppression of the scattering rate (by a factor
$\sim 1-\beta$) in the lab frame.}  To maintain $\tau_T < 1$, we require that
\begin{equation}
\Gamma_{\rm amb} > \Gamma_\tau(R) \equiv 
\Bigl[\tau_c^\prime(R)\,x_{\rm br}^{\alpha-2}\Bigr]^{1/(\alpha-1)}.
\end{equation}
The critical Lorentz factor $\Gamma_\tau$ can be expressed more
directly in terms of the initial bulk Lorentz factor of the ejecta,
after making use of eq. (\ref{eq:taucval}):
\begin{equation}
{\Gamma_\tau(R)\over\Gamma_{\rm ej}^0} \simeq
\left({R\over R_\tau}\right)^{-2/(\alpha-1)}.
\end{equation}
Note that $\Gamma_\tau$ coincides with $\Gamma_{\rm ej}^0$ 
at the scattering photosphere.
If $\Gamma_{\rm amb} < \Gamma_\tau$, then the $\gamma$-ray pulse is
spread out to a width
\begin{equation}
{\Delta t_{\rm spread}\over \Delta t} = {R\over 2\Gamma^2_{\rm amb} c\Delta t} 
\simeq \left({\Gamma_{\rm amb}\over \Gamma_\tau}\right)^{-2}\,
\left({R\over R_\tau}\right)^{(\alpha+3)/(\alpha-1)}.
\end{equation}
One finds that $\Delta t_{\rm spread} > \Delta t$ at $R \geq R_\tau$.

The kinetic energy of the accelerated medium moving at 
Lorentz factor $\Gamma_\tau$ (within a shell of thickness equal to the
acceleration length ${2\over 3}\Gamma_\tau^2c\Delta t$; eq. 
[\ref{eq:thinshell}]) can be no larger than the energy of the 
$\gamma$-ray pulse above a frequency $x \sim \Gamma_\tau$, 
\begin{equation}
\Gamma_\tau n_{p\,0}m_pc^2\,\left(1+2{n_{e^+}\over n_p}{m_e\over m_p}\right)
{2\over 3}\Gamma_\tau^2c\Delta t
\leq \left({F_{x_{\rm br}} x_{\rm br}\over c}\right)\,
\left({\Gamma_\tau\over x_{\rm br}}\right)^{1-\alpha}\,c\Delta t.
\end{equation}
This yields the upper bound
\begin{equation}
\rho_0 < {3x_{\rm br}^{(5-2\alpha)/(\alpha-1)}\over 
2(\tau_c^\prime)^{3/(\alpha-1)}}\left({m_e\over \sigma_T
c\Delta t}\right)\,\left(1+2{n_{e^+}\over n_p}{m_e\over m_p}\right)^{-1}
\propto \left({R\over R_\tau}\right)^{6/(\alpha-1)}\label{eq:rhozero}.
\end{equation}
The density in the above equation is higher than that encountered in the 
diffuse ISM, and therefore does not provide any constraint on GRB models 
involving the merging of compact objects. 
We can use this expression, however, to set limits on massive stars as 
likely GRB progenitors, as in this case the explosion occurs in a dense 
pre-burst stellar wind. Assuming a steady mass loss from such a massive 
progenitor at a rate $\dot M$ and constant speed $V$, the ambient 
density is 
\begin{equation}
\rho_0(R) = {\dot M\over 4\pi R^2 V}.
\end{equation}
Substituting this expression into eq. (\ref{eq:rhozero}), and making use 
of eqs. (\ref{eq:taucval}) and (\ref{eq:delt}) to express $R_\tau$
in terms of $\Delta t$, $\tau_c^{\prime\,0}$ and $x_{\rm br}$, we have
\begin{equation}
\dot M < 24\pi f_h \,
\Bigl(x_{\rm br}^{2\alpha-3}\,\tau_c^{\prime\,0}\Bigr)^{1/(\alpha-1)}\,
{m_ecV\Delta t\over \sigma_T}\,
\left({R\over R_\tau}\right)^{(2\alpha+4)/(\alpha-1)}.
\end{equation}
Setting $\Gamma_{\rm amb} = \Gamma_\tau$ and
eliminating $R$ , one deduces that the amount of pre-burst mass loss
needed to induce a certain spreading of the pulse is
\begin{equation}\label{eq:dtspread}
{\dot M\over \dot M_{\Delta t}} = 
\left({\Delta t_{\rm spread}\over\Delta t}\right)^{(2\alpha+4)/(3+\alpha)}.
\end{equation}
where
$$
\dot M_{\Delta t} \equiv 24\pi f_h\,
x_{\rm br}\,\Gamma_{\rm ej}^0\,\left({m_ecV\Delta t\over \sigma_T}\right)
\;\;\;\;\;\;\;\;\;\;\;\;\;\;\;\;\;\;\;\;\;\;\;
\;\;\;\;\;\;\;\;\;\;\;\;\;\;\;\;\;\;\;\;\;\;\;\;\;\;\;\;\;\;\;\;\;\;
\;\;\;\;\;\;\;\;\;\;\;\;\;\;\;\;
$$
\begin{equation}
= 1.5\times 10^{-5}\,x_{\rm br}f_h\,
\left({\Gamma_{\rm ej}^0\over 300}\right)\,
\left({\Delta t\over 10~{\rm s}}\right)\,
\left({V\over 1000~{\rm km~s^{-1}}}\right)\;\;\;\;\;\;{\rm M_\odot~yr^{-1}}.
\end{equation}
Notice that both the width $\Delta t_{\rm spread}$ of the spread pulse
(eq. [\ref{eq:delt}]) and $\dot M_{\Delta t}$ grow with radius $R$.  
Mass loss rates close to the critical value  $\dot M_{\Delta t}$ only
broaden the $\gamma$-ray pulse near the scattering photosphere, and do not
broaden it by much.  Higher mass loss rates cause further broadening
at larger radii.

The presence of a dense medium surrounding the GRB source
also limits the bulk Lorentz factor $\Gamma_{\rm ej}^0$ 
of the ejecta at the Thomson photosphere.  Hence the
width $\Delta t$ of the emitted $\gamma$-ray pulse (before
scattering in the ambient medium) cannot be made arbitrarily small.
An upper bound to $\Gamma_{\rm ej}^0$ is obtained by balancing
the energy of the ejecta with the kinetic energy of the 
baryons accumulating behind the shock:
\begin{equation}
{1\over \varepsilon_{\rm rad}}\,
{\partial E\over\partial\Omega} \geq {(\Gamma_{\rm ej}^0)^2\over 3-\delta}
\rho_0(R_\tau)R_\tau^3 c^2.
\end{equation}
Here $\varepsilon_{\rm rad}$ is the fraction of the kinetic energy
of the ejecta converted to $\gamma$-rays;  this expression is meant to
be evaluated at the scattering photosphere of the outflow, and so is missing
the factor of $\Gamma_{\rm amb}^{-1}$ present in eq. [\ref{eq:ehadron}].
Combining this equation with (\ref{eq:taucval}) and (\ref{eq:tauerel}),
one finds
\begin{equation}
\Gamma_{\rm ej}^0 \leq 
\left({4\pi V\over\varepsilon_{\rm rad}\dot Mc^2}\right)^{2/(5-\alpha)}\,
\left({\partial E\over\partial\Omega}\,
{m_ec^2\over\sigma_T}x_{\rm br}^{2-\alpha}\right)^{1/(5-\alpha)},
\end{equation}
or
\begin{equation}
\Gamma_{\rm ej}^0 \leq 36\,\varepsilon_{\rm rad}^{-2/3}\,
\left({\dot M\over 10^{-5}\,M_\odot~{\rm yr}^{-1}}\right)^{-2/3}\,
\left({V\over 1000~{\rm km~s^{-1}}}\right)^{2/3}\,
\left[{4\pi(\partial E/\partial\Omega)\over 10^{53}~{\rm erg}}\right]^{1/3}
\end{equation}
for a hard spectrum $\alpha = 2$. 
The width of the broadened pulse becomes
\begin{equation}
\Delta t \geq {R_\tau\over 2(\Gamma_{\rm ej}^0)^2c} \geq
{1\over 2c}\left({\varepsilon_{\rm rad}\dot M c^2\over 
4\pi V}\right)^{(3+\alpha)/(5-\alpha)}\,
\left({\partial E\over \partial\Omega}\right)^{(1-\alpha)/(5-\alpha)}\,
\left({m_ec^2\over\sigma_T}x_{\rm br}^{2-\alpha}\right)^{-4/(5-\alpha)}.
\end{equation}
Evaluating this expression for $\alpha = 2$, we deduce 
\begin{equation}\label{deltone}
\Delta t \geq 170\,\varepsilon_{\rm rad}^{5/3}\,
\left({\dot M\over 10^{-5}\,M_\odot~{\rm yr}^{-1}}\right)^{5/3}\,
\left({V\over 1000~{\rm km~s^{-1}}}\right)^{-5/3}\,
\left[{4\pi(\partial E/\partial\Omega)\over 10^{53}~{\rm erg}}\right]^{-1/3}
\;\;\;\;{\rm s}.
\end{equation}
We can combine this last result with eq. (\ref{eq:dtspread}) to obtain
a very similar bound on the width of the scattered $\gamma$-ray pulse:
\begin{equation}\label{delttwo}
{\Delta t_{\rm spread}\over\Delta t} = \left({1\over 
3f_h x_{\rm br}\varepsilon_{\rm rad}}\right)^{(\alpha+3)/(2\alpha+4)}\,
\Gamma_{\rm ej}^{(\alpha+3)(\alpha-2)/(2\alpha+4)(5-\alpha)}.
\end{equation}
Note that the width of the scattered pulse is independent of 
$\Gamma_{\rm ej}$ for a hard spectrum $\alpha = 2$, 
$\Delta t_{\rm spread}/\Delta t  
= 2.2\,(f_h x_{\rm br}\varepsilon_{\rm rad}/0.1)^{-5/8}$.  It is much
larger than $\Delta t$ for softer spectra, e.g.,
$\Delta t_{\rm spread}/\Delta t$ $= 2.1\,\Gamma_{\rm ej}^{3/10}\,
(f_h x_{\rm br}\varepsilon_{\rm rad}/0.1)^{-3/5}$ for $\alpha = 3$.
Because the softer spectrum creates fewer pairs, a larger
external mass density is required to induce runaway pair creation
for a fixed shock energy.

{\it This calculation sets significant constraints on GRB models
involving the delayed explosions of massive Wolf-Rayet stars}
(Woosley 1993; MacFadyen \& Woosley 1998), whose winds are 
characterized by mass loss rates $\dot M\approx 10^{-5}-10^{-4}\,M_\odot$
yr$^{-1}$ and velocities of 1000--2500 km s$^{-1}$ (Willis 1991).   
Hard bursts composed of peaks narrower than $\sim 10$ sec must
either form in a more rarefied environment, or have a low radiative efficiency
$\varepsilon_{\rm rad}$.  The possibility remains that the overall duration
of some $\gamma$-ray bursts is determined by the processes of pair-loading
and radiative acceleration in the ambient medium.  We note finally that the
constraint (\ref{deltone}), (\ref{delttwo}) is weakened if the $\gamma$-ray
source function is not a rigid power-law, but has a thermal cutoff below a
bulk frame energy $\sim m_ec^2$.

\section{Discussion}

We have calculated the deposition of radiative momentum into an 
(initially) static medium by the intense $\gamma$-ray pulse emanating from
a relativistic fireball (eqs. [\ref{eq:accone}] and [\ref{eq:acctwo}]).  
This process loads the accelerated medium
with pairs, which come to dominate its inertia (eq. \ref{eq:pairint});
and greatly increases the radiative efficiency of the forward shock
wave that bounds the fireball material.  Given this high radiative efficiency,
the ambient medium will attain a bulk Lorentz factor close to the 
Lorentz factor of the fireball material near the $\gamma$-ray photosphere,
decreasing as $R^{-2}$ at larger radius (eq. [\ref{eq:gamamb}]).
As a result, the rate of deceleration of the ejecta is slowed
(eq. [\ref{eq:gamej}]), and
the relation between the time-index of the synchrotron emission
and the synchrotron spectral index (eqs. [\ref{eq:betval}] and
[\ref{eq:betvalb}]) is modified from the relation previously
obtained for a static ambient medium. 

These physical processes have important implications for the radiative
physics of the forward shock, aside from its overall radiative efficiency.  
Anisotropy in the $\gamma$-ray flux (on an angular scale $\Gamma^{-1}$
or larger) will induce shearing motions in the external medium that can
strongly amplify any seed magnetic field before the shock hits it 
(\S \ref{magtangle}).  An even more important effect is that the minimum
Lorentz  factor of the shocked pairs is pushed down to $\gamma_e \sim 1$ in
the bulk frame, so that the corresponding synchrotron break frequency lies
at optical-UV near the $e^\pm$ scattering photosphere.  This may point
to a hybrid radiative mechanism (Thompson 1997) in which hard X-ray photons
advected out with the fireball are the principal seeds for
Comptonization by non-thermal pairs and bulk fluid motions.

\acknowledgments
Support for this work was provided by the Sloan foundation (C. T.), 
and by the NSF through grant PHY94-07194 (P. M.).  We thank A.M.
Beloborodov for discussions, and the referee for suggesting a number of
improvements to the presentation.

\references

Baring, M.G. \& Harding, A.K. 1997, \apj, 491, 663

Beloborodov, A.M. 1999, \mnras, 305, 181

Crider, A., Liang, E. P., Smith, I. A., Preece, R. D., Briggs,
M. S., Pendleton, G. N., Paciesas, W. S., Band, D. L., \& Matteson,
J. L. 1997, \apjl, 497, L39

%Galama, T. J., Wijers, R. A. M. J., Bremer, M., Groot, P. J., Strom, R. G.,
%Kouveliotou, C., \& Van Paradijs, J. 1998, \apjl, 500, L97

Jauch, J. M. \& Rohrlich, F. 1976, The Theory of Photons and Electrons 
(New York: Springer-Verlag) 

Katz, J. \& Piran T. 1997, ApJ, 490, 772

MacFadyen, A. \& Woosley, S. E. 1998, preprint (astro-ph/9810274)

Madau, P. \& Thompson, C. 1999, Ap J, in press (paper I)

Medvedev, M.V., \& Loeb, A. 1999, preprint ({\tt astro-ph/9904363})

M\'esz\'aros, P., Rees, M. J., \& Papathanassiou, H. 1994, \apj, 432, 181

M\'esz\'aros, P., \& Rees, M. J. 1997, \apj, 476, 232

M\'esz\'aros, P. Rees, M. J., \& Wijers, R. A. M. J. 1998, \apj, 499, 301

Noerdlinger, P.~D. 1974, \apj, 192, 529

O'Dell, S.~L. 1981, \apj, 243, L147

Phinney, E.~S. 1982, \mnras, 198, 1109

Sari, R., Piran, T., \& Narayan, R. 1998, \apj, 497, L17

Thompson, C. 1994, \mnras, 270, 480

Thompson, C. 1997, in Relativistic Jets in AGN, ed. M. Ostrowski, M.
Sikora, G. Madejski, \& M. Begelman, Krakow, p. 63

Waxman, E. 1997, \apjl, 485, L5

Wijers, R. A. M. J., Rees, M.J., \& M\'esz\'aros, P. 1997, \mnras, 288, L51

Willis, A. J. 1991, in Wolf-Rayet Stars and Interrelations with Other 
Massive Stars in Galaxies, ed. K. A. van der Hucht \& B. Hidayat
(Dordrecht: Kluwer), 256

Woosley, S. E. 1993, \apj, 405, 273

\vfill\eject

\begin{figure}
\plotone{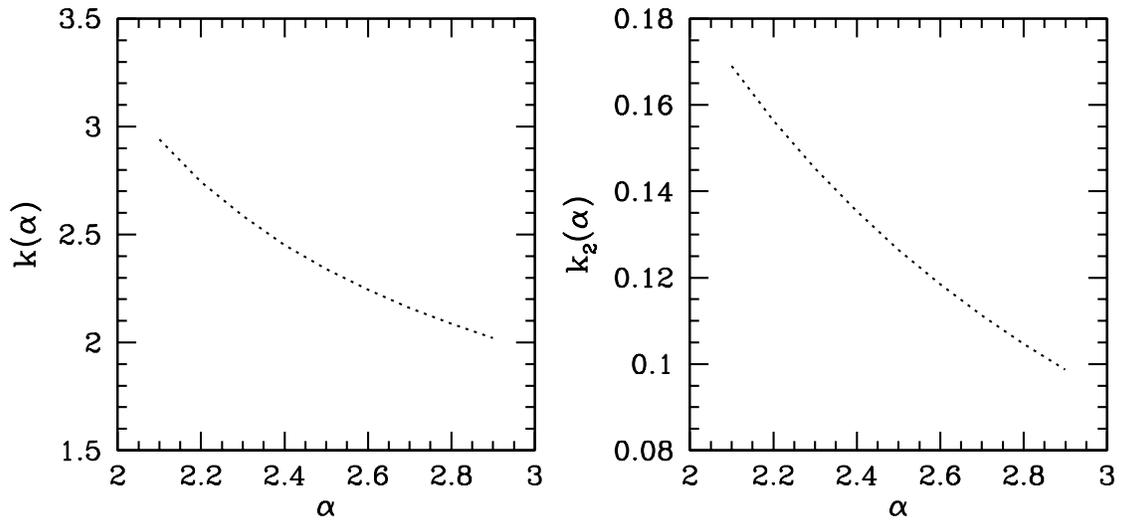}
\caption{{\it Left:} The function $k(\alpha)$ (eq. [\ref{eq:kdef}]),
plotted against spectral index $\alpha$.  {\it Right:} The function 
$k_2(\alpha)$ (eq. [\ref{taugmin}]).}
\label{fig1}
\end{figure}

\end{document}